\documentclass[12pt]{article}

\usepackage{amsmath}	
\usepackage{amssymb}	

\begin{document}

	\title{Zitterbewegung and a formulation for quantum mechanics}
	\author{G. S. Burra\\
		Adj. Professor, University of Udine, 201, Via delle Scienze\\
		33100, Udine, Italy}
	\date{}
	\maketitle
	\begin{abstract}
		In this short note, we try to show that inside a vortex like region, 
		like a 
		black hole one can observe superluminosity which yields some 
		interesting 
		results. Also, we consider the zitterbewegung fluctuations to obtain an 
		interpretation of quantum mechanics. We argue that this gives us a 
		method to 
		extract dark energy.
	\end{abstract}
	
	\section{Introduction}
	As is well known, the phenomenon of zitterbewegung has been studied by many 
	eminent authors like Schrodinger \cite{Erwin} and Dirac \cite{Dirac}. 
	Zitterbewegung is a type of rapid fluctuation that is attributed to the 
	interference between positive and negative energy states. Dirac pointed out 
	that the rapidly oscillating terms that arise on account of zitterbewegung 
	is averaged out when we take a measurement remembering that no measurement 
	is instantaneous.\\
	Interestingly, Hestenes \cite{Hestenes,David} too sought to explain quantum 
	mechanics through the phenomenon of zitterbewegung. Besides, the authors of 
	this paper have also studied the aforesaid phenomenon quite extensively in 
	the past few years \cite{bgs1,bgs2,bgs3,bgs4,abhi1,abhi2}. However, in this 
	paper we shall derive some interesting results that substantiate that the 
	Compton scale and zitterbewegung can delineate several phenomena.

	\section{Theory}
	Let us consider an object, such as a black hole or some particle, that is 
	spiralling like a vortex. In that case, we know that the circulation is 
	given by\\
	\[\Gamma = \oint_{C} v {\rm d}r = \frac{2\pi \hbar n}{m}\]\\
	where, $C$ denotes the contour of the vortex, $n$ is the number of turns 
	and $m$ is the mass of the circulating particle. Now, using the relation 
	derived by the author Sidharth \cite{sid3} and using the theory of vortices 
	\cite{Donnelly}\\
	\[v = \frac{D}{2\pi r}\]\\
	and the diffusion constant being given as\\
	\[D = \frac{\hbar}{m}\]
	we obtain from the integral\\
	\begin{equation}
		\Gamma = \frac{\hbar}{2\pi m}\oint_{l} \frac{{\rm d}r}{r} = \frac{2\pi 
		\hbar n}{m}
	\end{equation}
	Integrating this we derive the relation\\
	\begin{equation}
		r = e^{4\pi^{2}n}
	\end{equation}
	which shows that in case of vortices the radius of a particular vortex 
	increases with the number of turns, which is expected. One may wonder how 
	superluminosity, forbidden by special relativity and causality, can arise? 
	As we will see shortly that such ideas of superluminal velocities being 
	forbidden etc., arise above the Compton scale or more precisely outside the 
	zitterbewegung region belonging to the domain of usual physics. Now, 
	suppose the velocity equals that of the velocity of light. Then, we have 
	the circulation as\\
	\[\Gamma = \oint_{l} c {\rm d}r = \frac{2\pi \hbar n}{m}\]\\
	This implies that\\
	\[r = \frac{2\pi \hbar n}{mc}\]\\
	Thus, we have finally\\
	\begin{equation}
		r = 2\pi nl_{c} \label{1}
	\end{equation}
	where, $l_{c} = \frac{\hbar}{mc}$ is the Compton wavelength of the 
	circulating particle or object. Essentially, we have quantized the radius. 
	Now, it is known that the Langevin equation in the absence of external 
	forces is given as \cite{sid3}\\
	\[m\frac{{\rm d}v}{{\rm d}t} = -\alpha v + F^{\prime}(t)\]\\
	where, $\alpha$ is the coefficient of viscosity given by Stokes' law as\\
	\[\alpha = 6\pi \eta a\]\\
	Here, $a$ is the radius of the sphere under consideration. So, when we have 
	a cutoff time $\tau$, we get\\
	\[\tau \approx \frac{ma^{2}}{mca} = \frac{a}{c}\]\\
	Suppose, within the limit of the cut off time the radius of the sphere 
	under consideration coincides with that of the Compton length, then\\
	\[\tau \approx \frac{l_{c}}{c}\]\\
	Thus, from (\ref{1}) we have\\
	\[r = 2\pi nc\tau\]\\
	which finally yields\\
	\begin{equation}
		v_{c} = 2\pi nc \label{2}
	\end{equation}
	where, $v_{c} = \frac{r}{\tau}$ can be defined as the velocity in the 
	Compton scale. It is obvious from equation (\ref{2}) that the velocity 
	$v_{c}$ is greater than the velocity of light. This means that in the 
	Compton scale when a cut off time arises from considerations of stochastic 
	behaviour and the Langevin equation one derives superluminal velocities. 
	This is very interesting in the sense that in case of minimum spacetime 
	intervals one can expect phenomena that are different from usual Physics 
	that one observes above the Compton scale. It has also been pointed out by 
	Joos \cite{Joos} that from the Langevin like equation one can derive a 
	cutoff time that can define a limit for the Compton scale and the 
	phenomenon of zitterbewegung. Interestingly, when there is no cutoff time 
	then we get \cite{Reif}\\
	\[r = D\sqrt{t}\]\\
	where, $D$ is the diffusion constant. It is known that from the last 
	equation one can go to Nelson's derivation of the non-relativistic 
	Schrodinger equation \cite{sid3}. This implies that the cutoff time 
	provides a threshold limit beyond which Compton scale phenomena are not 
	observed or essentially, such stochastic dynamics are averaged out. 
	However, within the threshold limit we obtain results that are in stark 
	contrast with known physics. This is a key feature of the Compton scale 
	that has been discussed in several of our papers.\\

	\section{Consequence of a superluminal velocity}
	In the previous section we have obtained a novel result concerning the 
	velocity of the circulating particle. We have seen that considering a 
	vortex like motion of a particle in the Compton length one derives the 
	velocity of the particle to be greater than the velocity of light.\\
	Now, Sidharth had derived previously \cite{sid4} that for a Kerr-Newman 
	type charged black hole one would have an Aharonov-bohm type effect, on 
	account of the vector potential $\vec{A}$ which would give a shift for the 
	magnetic field as\\
	\[\Delta\delta_{B} = \frac{e}{\hbar}\oint \vec{A}.{\rm d}\vec{s}\]\\
	and the shift due to the electric charge would be\\
	\[\Delta\delta_{E} = -\frac{e}{\hbar}\oint A_{0}.{\rm d}t\]\\
	where, $\vec{A} \approx \frac{1}{c}A_{0}$. Using this relation it is easy 
	to deduce that the magnetic effect is $\sim \frac{v}{c}$ times the electric 
	effect. Now, when $v < c$ we have the electric effect to be stronger than 
	the magnetic effect. But, from our derivation in the previous section it is 
	obvious that the magnetic effect is stronger than the electric effect. This 
	also may imply that there is an extra magnetic effect arising from the 
	vortex like circulation of the particle. In case of a Kerr-Newman black 
	hole the magnetic field is given by\\
	\[B_{r} = \frac{2ea}{r^{3}}\cos\Theta\]\\
	\[B_{\Theta} = \frac{ea\sin\Theta}{r^{3}} + O(\frac{1}{r^{4}})\]\\
	\[B_{\phi} = 0\]\\
	Essentially, a short-ranged magnetic field that emerges due to the terms 
	$\sim \frac{1}{r^{4}}$ could result in the extra magnetic field that we 
	mentioned earlier. It is interesting to find that consideration of the 
	Compton region brings about such interesting results and phenomena. Now, we 
	would like to delve into another interesting aspect that stems from the 
	consideration of Compton scale. The author Sidharth had derived the quantum 
	mechanical spin from within the Compton scale \cite{sid5}. We would see 
	that from the vortex like consideration we obtain the same result.\\
	
	\section{The quantum mechanical spin}
	Let us consider equation (\ref{1}) once again. Multiplying both sides by 
	the momentum of the particle we have\\
	\[r\times p = 2\pi nl_{c}\times p\]\\
	Now, if the De Broglie wavelength of the particle with vortex features is 
	the Compton wavelength then we have\\
	\[p = \frac{\hbar}{l_{c}}\]\\
	Thus, we get\\
	\begin{equation}
		r\times p = 2\pi n\hbar
	\end{equation}
	Now, we know that the spin of a particle is given as\\
	\[S = \frac{\hbar}{2}\sqrt{n(n + 2)}\]\\
	where, $n$ is some positive integer. Then, considering the following 
	approximation\\
	\[4\pi n \approx \sqrt{N(N + 2)}\]\\
	for some positive integer $N$, we have the spin for the circulating 
	particle to be\\
	\begin{equation}
		r\times p = S \approx \frac{\hbar}{2}\sqrt{N(N + 2)} \label{3}
	\end{equation}
	Therefore, as we can see, we have obtained the quantum mechanical spin from 
	the consideration of the Compton scale and the vortex like feature of the 
	particle. This could be looked upon as the foundation of quantum 
	mechanics.\\
	
	\section{Discussions}
	In terms of positive and negative energy solutions we have\\
	\begin{equation}
		\psi_{p}
		=
		e^{\frac{i}{\hbar}Et}
		\begin{bmatrix}
			0 \\
			0 \\
			1 \\
			0
		\end{bmatrix}
		or,~
		e^{\frac{i}{\hbar}Et}
		\begin{bmatrix}
			0 \\
			0 \\
			0 \\
			1
		\end{bmatrix}
	\end{equation}
	\begin{equation}
		\psi_{n}
		=
		e^{-\frac{i}{\hbar}Et}
		\begin{bmatrix}
			1 \\
			0 \\
			0 \\
			0
		\end{bmatrix}
		or,~
		e^{-\frac{i}{\hbar}Et}
		\begin{bmatrix}
			0 \\
			1 \\
			0 \\
			0
		\end{bmatrix}
	\end{equation}
	From this Sidharth had derived the following relation \cite{sid5}\\
	\begin{equation}
		|\psi_{p} + \psi_{n}|^{2} = |\psi_{p}|^{2} + |\psi_{n}|^{2} + 
		(\psi_{p}\psi_{n}^{*} + \psi_{n}\psi_{p}^{*}) \label{a}
	\end{equation}
	It is interesting to note that the third term denotes that the positive and 
	negative energy solutions interact with each other and this happens in the 
	zitterbewegung region. So, basically, in this region classical physics gets 
	molded into a quantum mechanical version on account of the fluctuations 
	present. Outside this region, only the first two terms of the last equation 
	exist and thereby we observe normal physical phenomena that includes the 
	averaging out of the third term.\\
	We would like to add that the positive and negative domains are represented 
	by positive and negative energies insinuating forward and backward time 
	flow. In general, it is not possible to have both these domains coexisting. 
	However, it has been shown in great detail \cite{sidharth} that this is 
	represented within the Compton scale by a two Wiener process leading to 
	interesting results. It is worth mentioning that Newton and Wigner also had 
	argued that there is a region outside which we get the usual physical 
	phenomena.\\
	Now, in accordance with Wilson's renormalization group and one of our 
	previous papers we can consider localized wavepackets such that 
	\cite{Wilson1,wil,Wilson2,sid6}\\
	\begin{equation}
		\psi(x) = \psi_{m}(x) + \sum_{n}r_{n}\phi_{n}(x)
	\end{equation}
	where the modified wavefunction is $\psi_{m}(x)$ and each wavefunction 
	$\phi_{n}(x)$ fills a unit volume in the phase space. Here, the integration 
	is performed upon the coefficients $r_{n}$. Thus, considering Ito's lemma 
	\cite{Ito1,Ito2} as\\
	\[{\rm d}\psi(x_{t}) = \psi^{\prime}(x_{t}){\rm d}x_{t} + 
	\frac{1}{2}\psi^{\prime\prime}(x_{t})\sigma^{2}_{t}{\rm d}t\]\\
	we have\\
	\begin{equation}
		{\rm d}\psi_{m}(x_{t}) = \{\psi^{\prime}_{m}(x_{t}) + 
		\sum_{n}r_{n}\phi^{\prime}_{n}(x_{t})\}{\rm d}x_{t} + 
		\frac{1}{2}\{\psi^{\prime\prime}_{m}(x_{t}) + 
		\sum_{n}r_{n}\phi^{\prime\prime}_{n}(x_{t})\}\sigma^{2}_{t}{\rm d}t - 
		\sum_{n}{\rm d}\{r_{n}\phi_{n}(x_{t})\}
	\end{equation}
	Interestingly, visualizing $\phi(x)$ as a step function we were able to 
	derive the following relation
	\begin{equation}
		\phi^{\prime\prime}(x_{t}) =  \delta^{\prime}(l)\frac{{\rm d}l}{{\rm 
		d}x_{t}} + \delta(l)\frac{{\rm d^{2}}l}{{\rm d}x_{t}^{2}} \label{4}
	\end{equation}
	Again, we know that in electromagnetism \cite{Jackson,Zangwill}, the
	gradient of the delta function represents a point magnetic dipole
	situated at the origin and that the function itself represents a
	point charge \cite{Glen}, owing to it's distributional property. Thus, we 
	can see that the electric charge emerges from the Compton scale when one 
	considers the wavefunction $\phi(x_{t})$ to describe the phase space of 
	unit volume. This phase space can essentially be looked upon as a cloud 
	like space with fluctuations going on inside it. The properties of this 
	cloud like space is entirely embedded in the wavefunction $\phi(x_{t})$. 
	From equation (\ref{4}) we derive the electric charge that can be 
	attributed to the collective fluctuations or zitterbewegung going on inside 
	the phase space. When an observation is made upon this cloud like space the 
	electron is created instantaneously.\\
	We would also like to infer that these fluctuations are on account of the 
	quantum vacuum that exists in the phase space. Considering the methodology 
	of a quantized Klein-Gordon field in the vacuum state, one can calculate 
	the probability that the configuration of this cloud like space is given by 
	$\phi(x_{t})$.\\
	\begin{equation}
		\rho_{0}[\phi(x_{t})] = \exp[-\frac{1}{\hbar}\int 
		\frac{1}{(2\pi)^{3}}\phi^{*}(k_{t})\phi(k_{t})\sqrt{|k|^{2} + 
		m^{2}}{\rm d}^{3}k_{t}] \label{5}
	\end{equation}
	where, $\phi(k_{t})$ is the Fourier transform of $\phi(x_{t})$ which 
	essentially characterizes the electron. Incidentally, this implies that the 
	electron is the Fourier transform of the cloud like structured phase space. 
	Inasmuch, the cloud like space is connected to the point electron through a 
	Fourier transform when the region is observed. Physically, one can argue 
	that the very act of observation of the quantum vacuum leads to the 
	detection of the electron and mathematically is represented by a Fourier 
	transform given by equations of type (\ref{5}).\\
	Also, we had seen earlier that within the Compton scale or more precisely 
	the zitterbewegung region we have "unphysical effects" like superluminosity 
	and a breakdown of causality \cite{abhi3,abhi4}.
	But, these are tweaked away once we return to the physical region. The 
	whole 
	point is that these ideas are based on notions of space time points which 
	as 
	noted by authors like Rohrlich \cite{Rohr1,Rohr2} is an oxymoron.
	\section{Remark} We have seen that the electro magnetic tensor and current 
	vector have been obtained, however weak they maybe. This is given by 
	\begin{equation}\label{emf}
		F_{\mu\nu}^{\prime} = F_{\mu\nu} - \epsilon (F_{\mu\nu})_{0}
	\end{equation}
	and
	\begin{equation}\label{current}
		j_{\mu}^{\prime} = j_{\mu} - \epsilon (j_{\mu})_{0}
	\end{equation}
	where the subscript ``0'' referrs to the contribution of dark energy.
	This is an example of the real footprint of the ethereal dark energy. From 
	this 
	we should be able to extract energy. It’s a question of  technology.
	\section{Conclusions}
	Thus, from various points of view we come to the conclusion that the 
	zitterbewegung region is eliminated during observable physical processes.
	What we have argued is that classical physics represented by the first two 
	terms of equation (\ref{a}) where the positive and negative energy solution 
	stand decoupled into separate domains. But, on the contrary, it is the 
	interference of these two that gives rise to quantum mechanics.


\begin{thebibliography}{99}
		\bibitem{Erwin} E. Schrodinger, \emph{Report on the Quantum Dynamics of 
		the Electron}, Meeting of the Prussian Academy of Sciences. 
		Physico-mathematical class, 63-72, 1931.
		
		\bibitem{Dirac} P. Dirac, \emph{Principles of Quantum Mechanics}, 
		Oxford University Press, 1930.
		
		\bibitem{Hestenes} D. Hestenes, \emph{Quantum Mechanics from 
		Self-Interaction}, Foundations of Physics, Volume 15, Issue 1, pp 
		63-87, 1985.
		\bibitem{David} D. Hestenes, \emph{The Zitterbewegung Interpretation of 
		Quantum Mechanics}, Found. Physics., Vol. 20, No. 10, (1990) 1213-1232.
		\bibitem{bgs1} B.G. Sidharth, \emph{Revisiting Zitterbewegung}, IJTP, 
		Volume 48, Issue 2, pp 497-506, 2009.
		\bibitem{bgs2} B.G. Sidharth, \emph{ZPF, Zitterbewegung and Inertial 
		Mass} Int. J. Mod. Phys. E, 18, 1863, 2009.
		\bibitem{bgs3} B.G. Sidharth, \emph{The New Cosmos}, Chaos, Solitons 
		and Fractals, vol. 18, no. 1, pp. 197-201, 2003.
		\bibitem{bgs4} B.G. Sidharth, \emph{An Underpinning for Spacetime}, 
		Chaos, Solitons and Fractals 25, pp.965-968, 2005
		\bibitem{abhi1} B.G. Sidharth \& A. Das, \emph{The Zitterbewegung 
		Region}, IJMPA, Vol. 32, Nos. 19 \& 20 (2017) 1750117.
		\bibitem{abhi2} B.G. Sidharth \& A. Das, \emph{Comments on the paper 
		"The zitterbewegung region"}, IJMPA, Vol. 32, No. 28n29, 1750173 (2017).
		
		\bibitem{sid3} B. G. Sidharth, \emph{The Thermodynamic Universe}, pg 
		267 \& pg 57, World Scientific, 2008.
		
		\bibitem{Donnelly} R. J. Donnelly, \emph{Quantized Vortices in Helium 
		2}, pp. 1-41, Cambridge University Press, 1991.
		
		\bibitem{Joos} G. Joos, \emph{Theoretical Physics}, pp. 212, Blackie \& 
		Son Limited, London and Glasgow, 1951.
		
		\bibitem{Reif} F. Reif, \emph{Fundamentals of Statistical and Thermal 
		Physics}, McGraw-hill Book Co., Singapore, 1965.
		
		\bibitem{sid4} B. G. Sidharth, \emph{Chaotic Universe}, Nova Science, 
		New York, 2001.
		
		\bibitem{sid5} B. G. Sidharth, \emph{A brief note on "extension, spin 
		and non-commutativity"}, Foundations of Physics Letters, Volume 15, 
		Issue 5, pp 501-506, 2002.
		
		\bibitem{sidharth} B. G. Sidharth, \emph{The Universe of Fluctuations: 
		The Architecture of Spacetime and the Universe}, pp. 64, Springer, ISBN 
		978-1-4020-3786-3, 2005.
		
		\bibitem{Wilson1} K. G. Wilson, \emph{The Renormalization Group and 
		Critical Phenomena}, Reviews of Modern Physics, Vol. 55, No. 3, 1983.
		\bibitem{wil} K. G. Wilson, \emph{Renormalization group methods}, 
		Advances in Mathematics, Volume 16, Issue 2, May 1975, Pages 170-186
		\bibitem{Wilson2}  K. G. Wilson, and  J. Kogut, \emph{The 
		renormalization group and the $\epsilon$ expansion}, Physics Reports,
		Volume 12, Issue 2, 1974, Pages 75-199.
		
		\bibitem{sid6} B. G. Sidharth \& A. Das, \emph{A note on gravitation 
		and electromagnetism}, MPLA, Vol. 33, No. 13, 1850071 (2018).
		
		\bibitem{Ito1} K. Ito, \emph{Stochastic Integral}, Proc. Imperial Acad. 
		Tokyo 20, 519-524, 1944.
		\bibitem{Ito2} K. Ito, \emph{On stochastic differential equations}, 
		Memoirs, American Mathematical Society 4, 1-51, 1951.
		\bibitem{Jackson} J. D. Jackson, \emph{Classical Electrodynamics}, 
		Wiley Eastern Limited, pp. 180-188, 1975.
		\bibitem{Zangwill} A. Zangwill, \emph{Modern Electrodynamics}, 
		Cambridge University Press, pp. 343-344, 2013.
		\bibitem{Glen} G. S. Smith, \emph{An Introduction to Classical 
		Electromagnetic Radiation}, Cambridge University Press, pp. 331-335, 13 
		Aug 1997.
		\bibitem{abhi3} B.G. Sidharth \& A. Das, \emph{Space–time geometry and 
		the velocity of light}, IJMPA, Vol. 33, No. 30, 1850183 (2018).
		\bibitem{abhi4} B.G. Sidharth \& A. Das, \emph{A note on 
		superluminosity}, MPLA, Vol. 34, No. 22, 1950173 (2019).
		
		\bibitem{Rohr1} F. Rohrlich, \emph{Dynamics of a charged particle}, 
		Physical Review E, 77, 046609, 2008.
		\bibitem{Rohr2} F. Rohrlich, \emph{The dynamics of a charged sphere and 
		the electron}, American Journal of Physics, 65 (11), 1997.
		
	\end{thebibliography}
	\end{document}